\documentclass[
twocolumn,
aps,
prb,
reprint,
superscriptaddress,
amsmath,amssymb,
]{revtex4-1}
\usepackage[pdftex]{graphicx, color}
\usepackage[pdftex,
            colorlinks=true,
            citecolor=blue,
            linkcolor=blue]{hyperref}
\usepackage{braket,ulem}
\usepackage{times,multirow,amsfonts,bm,xspace,pifont}
\usepackage{soul}
\usepackage{here}
\begin{document}
\newcommand{\rr}{{\bm r}}
\newcommand{\q}{{\bm q}}
\renewcommand{\k}{{\bm k}}
\newcommand*\wien    {\textsc{wien}2k\xspace}
\newcommand*\textred[1]{\textcolor{red}{#1}}
\newcommand*\textblue[1]{\textcolor{blue}{#1}}
\newcommand{\ki}[1]{{\color{red}\st{#1}}}
\newcommand{\sgn}{\mathrm{sgn}\,}
\newcommand{\tr}{\mathrm{tr}\,}
\newcommand{\Tr}{\mathrm{Tr}\,}
\newcommand{\GL}{{\mathrm{GL}}}
\newcommand{\talpha}{{\tilde{\alpha}}}
\newcommand{\tbeta}{{\tilde{\beta}}}
\newcommand{\mathN}{{\mathcal{N}}}
\newcommand{\mathQ}{{\mathcal{Q}}}
\newcommand{\bv}{{\bar{v}}}
\newcommand{\bj}{{\bar{j}}}
\newcommand{\zero}{{(0)}}
\newcommand{\one}{{(1)}}
\newcommand{\two}{{(2)}}
\newcommand{\three}{{(3)}}
\newcommand{\four}{{(4)}}

\newcommand{\YY}[1]{\textcolor{blue}{#1}}
\newcommand{\AD}[1]{\textcolor{magenta}{#1}}
\newcommand{\ADS}[1]{\textcolor{magenta}{\sout{#1}}}

\graphicspath{{./fig_submit/}}

% Title of paper
\title{Orbital effect on intrinsic superconducting diode effect
%of helical superconductivity
%in $s$- and $d$-wave superconductors
}

\author{Kyohei Nakamura}
\email[]{nakamura.kyohei.84w@st.kyoto-u.ac.jp}
\affiliation{Department of Physics, Graduate School of Science, Kyoto University, Kyoto 606-8502, Japan}
% \author{Yuhei Ikeda}
% \affiliation{Department of Physics, Graduate School of Science, Kyoto University, Kyoto 606-8502, Japan}
\author{Akito Daido}
\affiliation{Department of Physics, Graduate School of Science, Kyoto University, Kyoto 606-8502, Japan}
\author{Youichi Yanase}
\affiliation{Department of Physics, Graduate School of Science, Kyoto University, Kyoto 606-8502, Japan}
% \affiliation{%
%   Institute for Molecular Science, Okazaki 444-8585, Japan
% }%
\date{\today}

\begin{abstract}
Much ink has recently been spilled on nonreciprocal phenomena in superconductors, especially the superconducting diode effect (SDE) characterized by the nonreciprocity of the critical current $\Delta J_{\rm c}$. 
Contrary to the fundamental and practical significance of the SDE, the precise underlying mechanism remains unclear.
In this paper, we investigate the impact of an orbital effect on the intrinsic SDE in a bilayer superconductor with Rashba spin-orbit coupling and an in-plane magnetic field.
We show that a small orbital effect leads to the sign reversal of $\Delta J_{\rm c}$ and a crossover of the helical superconducting state at a lower magnetic field than the monolayer superconductor.
On the other hand, a large orbital effect induces a decoupling transition, stabilizing a finite momentum Cooper pairing state called the orbital Fulde-Ferrell-Larkin-Ovchinnikov state, and results in the drastic change of the SDE. 
Owing to the orbital effect, the field dependence of the SDE may show oscillations several times. %and the Cooper pair momentum linearly increases 
%as increasing the magnetic field.
The results shed light on the mechanism of the SDE in atomically-thin multilayer superconductors.
\end{abstract}

\maketitle

\section{INTRODUCTION}
%Much ink has recently been spilled on 
Recent development in condensed matter physics has clarified the significance of nonreciprocal phenomena~\cite{Tokura2018-nb,Ideue2021-es}, which enable us to explore a fertile ground for quantum materials, such as the inversion and time-reversal symmetries, the spin-orbit coupling (SOC), the quantum geometry, and the strong correlation effect.
Nonreciprocity is characterized by the directional transport of particles or quasi-particles.
As for the nonreciprocal transport of electrons, the magnetochiral anisotropy (MCA), which refers to the directional electric resistance, has been observed in various noncentrosymmetric materials~\cite{Rikken2001-il,Krstic2002-vo,Pop2014-wn,Rikken2005-ew,Ideue2017-vp,Wakatsuki2018-ll,Hoshino2018-sa,Wakatsuki2017-dp,Qin2017-vd,Yasuda2019-jw,Itahashi2020-ef}.
In superconductors near the transition temperature, the superconducting fluctuation also induces the nonreciprocity of the electric conductivity, which is called the nonreciprocal paraconductivity, a kind of the MCA~\cite{Wakatsuki2017-dp,Qin2017-vd,Yasuda2019-jw,Itahashi2020-ef,Zhang2020-al}. 
Observation of the MCA is of prime importance in detecting an unconventional superconducting state, in particular, the spin-singlet and -triplet mixed superconducting state~\cite{Wakatsuki2018-ll} and the helical superconducting state~\cite{Daido_paraconductivity}.

Now for nonreciprocal phenomena in superconductors, the superconducting diode effect (SDE) has received considerable scholarly attention in recent years~\cite{Ando2020-om,Miyasaka2021-ly,Shin2021-je,Lin2021-xr,Bauriedl2022-nq,Narita2022-od,Yuan2022-pz,Daido2022-ox,He2022-px,Ilic2022-kh,Scammell2022-pv,Zhai2022-we,Karabassov2022-xu,Legg2022-ol,Lyu2021-sm,Dobrovolskiy2022-gk,Hou2022-hu,Sundaresh2022-aa,Hope2021-st,Zinkl2021-jl,Jiang2022-lr,Baumgartner2021-lg,Baumgartner2022-gi,Wu2022-ey,Diez-Merida2021-ih,Pal2021-df,Gupta2022-wn,Turini2022-hz,Hu2007-cu,Kopasov2021-xz,Misaki2021-wt,Halterman2022-lc,Zhang2021-ju,Davydova2022-yu,Souto2022-bb,Tanaka2022-qw,Zhang2022-lo,Suri2022-vl,Xie2022-of,Li2022-vm,Paolucci2023-gf,Ikeda2022-do,Chazono2022-pi,Kawarazaki2022-sd,Gutfreund2023-dv,Kealhofer2023-as,Kokkeler2022-fj,Trahms2023-sa,Ghosh2022-ht,Daido2022-sd,Mizuno2022-ht,Fu2021-co}.
The SDE refers to the directional critical current for the phase transition between normal and superconducting states and is characterized by the nonreciprocity of the critical current $\Delta J_{\rm c}$.
The SDE has been observed in various superconductors, such as the Nb/V/Ta superlattice and Rashba heterostructures~\cite{Ando2020-om,Miyasaka2021-ly,Narita2022-od,Kawarazaki2022-sd}, transition metal dichalcogenides NbSe$_2$~\cite{Shin2021-je,Bauriedl2022-nq}, twisted multilayer graphene~\cite{Lin2021-xr}, conventional superconductors~\cite{Lyu2021-sm,Hou2022-hu,Sundaresh2022-aa,Suri2022-vl,Gutfreund2023-dv}, and cuprate superconductors~\cite{Mizuno2022-ht}.
The Josephson junction is also the platform for the SDE, which is called the Josephson diode effect (JDE)~\cite{Baumgartner2021-lg,Baumgartner2022-gi,Wu2022-ey,Diez-Merida2021-ih,Pal2021-df,Gupta2022-wn,Turini2022-hz,Hu2007-cu,Kopasov2021-xz,Misaki2021-wt,Halterman2022-lc,Zhang2021-ju,Davydova2022-yu,Souto2022-bb,Tanaka2022-qw,Kokkeler2022-fj,Trahms2023-sa,Ghosh2022-ht}.
Observation of the SDE and JDE at high temperatures provides a path toward realizing the dissipationless electric circuits~\cite{Mizuno2022-ht,Ghosh2022-ht}.
Work is also currently underway to design a device with the SDE~\cite{Paolucci2023-gf}. 

To investigate the SDE, it is essential to ascertain what determines the critical current.
Generally speaking, the critical current is determined by the vortex mechanisms, such as the vortex surface barrier~\cite{Hope2021-st,Suri2022-vl,Hou2022-hu}, vortex ratchet~\cite{Li2022-vm}, and vortex-antivortex pairs~\cite{Jiang2022-lr}.
Because the dynamics of vortices depend on the experimental setup such as sample geometry and quality, the SDE derived from the vortex mechanisms is called the extrinsic SDE.
The extrinsic SDE, however, can not describe all the experimental results.
For example, the oscillation of $\Delta J_{\rm c}$ for a magnetic field~\cite{Kawarazaki2022-sd} has not been explained based on the extrinsic SDE.
Another approach to understanding the SDE is to focus on the depairing critical current, which reflects the intrinsic properties of superconductors. The SDE determined by the nonreciprocity of depairing critical current is called the intrinsic SDE~\cite{Yuan2022-pz,Daido2022-ox,He2022-px,Ilic2022-kh,Daido2022-sd,Ikeda2022-do}.
Theoretical studieds on the intrinsic SDE~\cite{Daido2022-ox,Ilic2022-kh,Daido2022-sd} have predicted the oscillation of $\Delta J_{\rm c}$ for a magnetic field, and revealed a close relationship between the SDE and a finite Cooper pair momentum state dubbed as helical superconductivity~\cite{Bauer2012-xi,Smidman2017-hb,Agterberg2003-jn,Barzykin2002-eh,Dimitrova2007-hp,Samokhin2008-nv,Yanase2008-yb,Michaeli2012-gl,Sekihara2013-dm,Houzet2015-iy,Dimitrova2003-mo,Kaur2005-jf,Agterberg2007-vl}.
The SDE is therefore expected to be valuable means to detect helical superconductivity, while it is challenging to identify the helical superconductivity with conventional experimental methods.
On the other hand, there is yet no firm evidence that the SDE is due to the intrinsic mechanism. 
From the above, contrary to the fundamental and practical significance of the SDE, there is a dearth of information about the precise underlying mechanism.
Further study is eagerly awaited to solve the problem.

%The bulk materials where the SDE is observed are the layered superconductors, where different layers are weakly coupled by Josephson coupling or van-der Waals interaction.
Many platforms of the SDE, including the Nb/V/Ta superlattice, are heterostructures where superconducting layers are weakly coupled. 
When the thickness of each layer is smaller than the coherence length, the Abrikosov vortex can not penetrate the layers under the in-plane magnetic field.
Hence, most previous studies have investigated the SDE by neglecting the orbital effect, assuming the monolayer limit.
However, the Josephson vortex can penetrate between the layers,
%and change the order parameter of multilayer heterostructures.
%and induce the orbital effect.
%In addition, it has been pointed out that 
and indeed, the thickness of layers in the Nb/V/Ta superlattice is not thin enough to completely neglect the Josephson vortex or the orbital effect~\cite{Fu2021-co,Ando2021_orbital}.
Because the Josephson vortex changes the order parameter of multilayer superconductors, an intrinsic property of superconductors, 
%it is expected to be particularly important for the intrinsic SDE. 
%results from the depairing mechanism.
it is naturally expected that the orbital effect plays an essential role in the SDE. However, a systematic understanding of the orbital effect on the SDE is still lacking.
This naturally motivates us to gain a decipherment of the orbital effect on the intrinsic SDE, which may shed light on the precise mechanism of the experimentally observed SDE.

In this paper, we investigate the impact of the orbital effect on the intrinsic SDE in a bilayer system, a minimal platform of multilayer superconductors.
In Sec.~\ref{sec:model}, we introduce a model of the SDE in bilayer $s$-wave superconductors with the Rashba-Zeeman effect.
The orbital effect caused by an in-plane magnetic field is built into the model.
In Sec.~\ref{sec:orbital effect on the SDE}, we analyze the data gathered and address each of the research questions in turn.
We show that a small orbital effect results in a crossover of helical superconductivity at a lower field than the monolayer system and a large orbital effect induces the decoupling of two layers resulting in the Josephson vortex state. These changes in the superconducting state alter the behaviors of the SDE and in turn, they can be detected by the SDE.
In Sec.~\ref{sec:summary and discussion}, we summarize the paper and discuss the experimental observations in the Nb/V/Ta superlattice. 
%and \YY{comment on recent work on the SDE and orbital Fulde-Ferrell-Larking-Ovchinnikov state~\cite{Xie2022-of}.} 

%%%%%%%%%%%%%%%%%%%%%%%%%%%%%%%%%%%%%%%%%%%%%%%%%%%%
\section{MODEL AND SETUP}
\label{sec:model}

In this section, we set up a model to investigate the orbital effect on the SDE.
We study a typical platform of the SDE, the $s$-wave superconductor with the Rashba and Zeeman effects~\cite{Daido2022-ox,Yuan2022-pz,Ilic2022-kh}, and extend the model to the bilayer structure:
%Following Ref.~\onlinecite{Daido2022-ox}, we treat two-dimensional $s$-wace superconductivity with Rashba-Zeeman effect and an in-plane magnetic field. 
%In the present work, furthermore, we introduce a model for a bilayer system so as to consider the orbital effect:
\begin{align}
\hat{H}\!\!&=\!\!\!\!\sum_{\bm k \sigma\sigma'm}\!\!\!\bigl[\xi (\bm{k}\!+\!\bm{p}_m)\delta_{\sigma\sigma'}
\!+\!\{\bm{g}(\bm{k}\!+\!\bm{p}_m)\!-\!\bm{h}\}\! \cdot \!\bm{\sigma}_{\sigma\sigma'}]
c_{\bm k \sigma m}^\dagger c_{\bm k \sigma' m} \notag 
\\
&\quad-\frac{U}{V}\sum_{\bm{k},\bm{k}',\bm{q},m}c^\dagger_{\bm{k}+\bm{q}\uparrow m}c^\dagger_{-\bm{k}\downarrow m}c_{-\bm{k}' \downarrow m}c_{\bm{k}'+\bm{q} \uparrow m}\notag \label{eq:interaction}%\\
\\
&\quad+t_{\perp}\sum_{\bm k \sigma \langle mm' \rangle}c^\dagger_{\bm k \sigma m}c_{\bm k \sigma m'} ,
\end{align}
where $\bm{k}$, $\sigma=\uparrow,\downarrow$ and $m=1,2$ are index of momentum, spin, and layer, respectively. 
In the first line of Eq.~\eqref{eq:interaction}, the normal-state Bloch Hamiltonian of each layer is given by 
\begin{align}
 H_N^m(\bm{k})=\xi(\bm{k}+\bm{p}_m)+(\bm{g}(\bm{k}+\bm{p}_m)-\bm{h})\cdot\bm{\sigma},  
\end{align}
which includes hopping energy and chemical potential
\begin{align}
\xi(\bm{k})=-2t(\cos k_x+\cos k_y)-\mu,
\end{align}
the Rashba SOC with
\begin{align}
    \bm{g}(\bm{k})=\alpha_{\rm g}(-\sin k_y,\sin k_x,0),
\end{align}
and the Zeeman term $\bm{h}\cdot\bm{\sigma}$.
Here, the magnetic field is applied along the $y$ direction, $\bm{h}=(0,\mu_{B}H,0)$.
The energy dispersion of the monolayer Hamiltonian $H_N^m(\bm{k})$ is obtained as
\begin{align}
    \xi_\chi(\bm{k},h)&=\xi(\bm{k})+\chi|\bm{g}(\bm{k})-h\hat{y}| \notag
    \\
    &\simeq\xi_\chi(\bm{k}-\delta\bm{q}_\chi(\bm{k},h),0),
\end{align}
where each band is labeled by the helicity $\chi=\pm1$, and the helicity-dependent momentum shift due to the Zeeman effect $h=\mu_{B}H$ is given by 
\begin{align}
    \delta\bm{q}_\chi(\bm{k},h)=\frac{\chi g_y(\bm{k})h}{|\bm{g}(\bm{k})|\cdot|\nabla_{\bm{k}}\xi_\chi(\bm{k},0)|^2}\nabla_{\bm{k}}\xi_\chi(\bm{k},0). \label{eq:momentum shift from para}
\end{align}
In addition, a layer-dependent momentum shift is caused by the orbital effect arising from the in-plane magnetic field. 
When the vector potential is chosen as $\bm{A}=(Hz,0,0)$, the momentum shift is obtained as
\begin{align}
    \bm{p}_m&=\frac{e}{\hbar}\bm{A}=\frac{eHD}{\hbar}\left(\frac{3}{2}-m\right)\hat{x} \notag
    \\
    &\equiv hd\left(\frac{3}{2}-m\right)\hat{x}. \label{eq:orbital effect}
\end{align}
Here, $e$, $\hbar$, and $D$ are the electron charge, Dirac constant, and interlayer distance, respectively.
The orbital effect is parameterized by $d=eD/\mu_B\hbar$.

In the second line of Eq.~\eqref{eq:interaction}, we adopt a pairing interaction for $s$-wave superconductivity and a finite momentum of Cooper pairs $\bm{q}$ which can be assumed to be $\bm{q}=q\hat{x}$ because of the $y$ mirror symmetry.
The system size is represented by $V=L_xL_y$, and we adopt $L_x=3000$ and $L_y=200$ for numerical calculations.
We choose $q$ to be compatible with the periodic boundary conditions, $q\in2\pi\mathbb{Z}/L_x$.
In the third line of Eq.~\eqref{eq:interaction}, $t_{\perp}$ is an inter-layer hopping and assumed to be much smaller than the in-plane hopping $t$.
The parameters are chosen as
%\begin{align}
$
(t,\mu,\alpha,U,t_{\perp})=(1,-1,0.3,1.5,0.1),
$ 
%\end{align}
unless mentioned otherwise, and give the transition temperature of superconductivity $T_{\rm c} \simeq 0.039$ at the zero magnetic field.
In particular, $T_{\rm c}$ is independent of $d$.

We calculate the depairing critical current by using the mean-field approximation. 
The gap equation for the order parameter on each layer is introduced by
\begin{align}
    \Delta_m(q)=-\frac{U}{V}\sum_{\bm{k}}\langle c_{-\bm{k}\downarrow m}c_{\bm{k}+q\hat{x}\uparrow m}\rangle . \label{eq:gap}
\end{align}
Then, the mean-field Hamiltonian $\hat{H}_{\mathrm{MF}}(q)$ and the Bogoliubov-de Gennes (BdG) Hamiltonian $\mathcal{H}_{q}(\bm{k})$ are obtained as
\begin{align}
    \hat{H}_{\mathrm{MF}}(q)&=\frac{1}{2}\sum_{\bm{k}}\Psi_{q}^{\dagger}(\bm{k})\mathcal{H}_{q}(\bm{k})\Psi_{q}(\bm{k}) \notag
    \\
    &\quad+\frac{1}{2}\sum_{\bm{k}\sigma m}[[H_N^m(\bm{k})]_{\sigma\sigma}+\Delta_m(q)^2/U] ,
    \\
    \notag
    \\
    \mathcal{H}_{q}(\bm{k})&=\begin{pmatrix}
        H_{q}^1(\bm{k})&\tau\\
        \tau&H_{q}^2(\bm{k}) 
    \end{pmatrix}
    ,
\end{align}
with the layer-dependent BdG Hamiltonian
\begin{align}
H_{q}^m(\bm{k})&=\begin{pmatrix}
    H_N^m(\bm{k}+q\hat{x})&\Delta_m(q)i\sigma_y \\
    -\Delta_m(q)i\sigma_y&-^{t}H_N^m(-\bm{k})
\end{pmatrix}
,
\end{align}
and the inter-layer hopping Hamiltonian
\begin{align}
\tau=\begin{pmatrix}
    t_{\perp}\bm{I}_2&\bm{0} \\
    \bm{0}&-t_{\perp}\bm{I}_2
\end{pmatrix}
.
\end{align}
The Nambu spinor is defined as
\begin{align}
\Psi_{q}^{\dagger}(\bm{k)}&=(\Psi_{q1}^\dagger(\bm{k}),\Psi_{q2}^\dagger(\bm{k})) ,
\end{align}
\begin{align}
\Psi_{qm}^\dagger(\bm{k})=(c_{\bm{k}+q\hat{x}\uparrow m}^\dagger,c_{\bm{k}+q\hat{x}\downarrow m}^\dagger ,c_{-\bm{k}\uparrow m},c_{-\bm{k}\downarrow m}). 
\end{align}

\begin{figure*}[htbp]
    \centering  
      \begin{tabular}{lcl}
      (a)&$\quad$&(b)\\
          \includegraphics[width=0.45\textwidth,height=0.25\textheight]{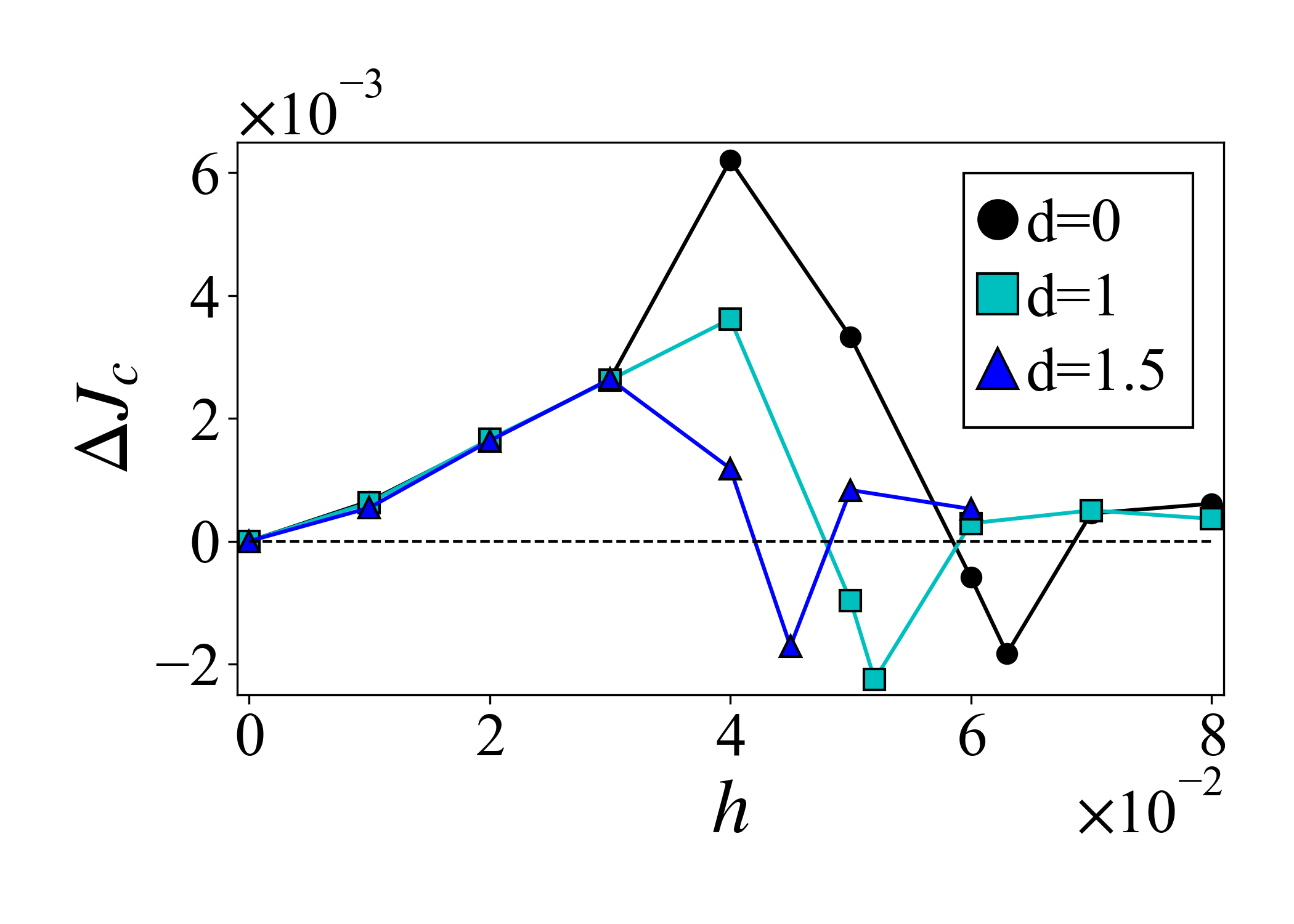}
      &&\includegraphics[width=0.45\textwidth,height=0.25\textheight]{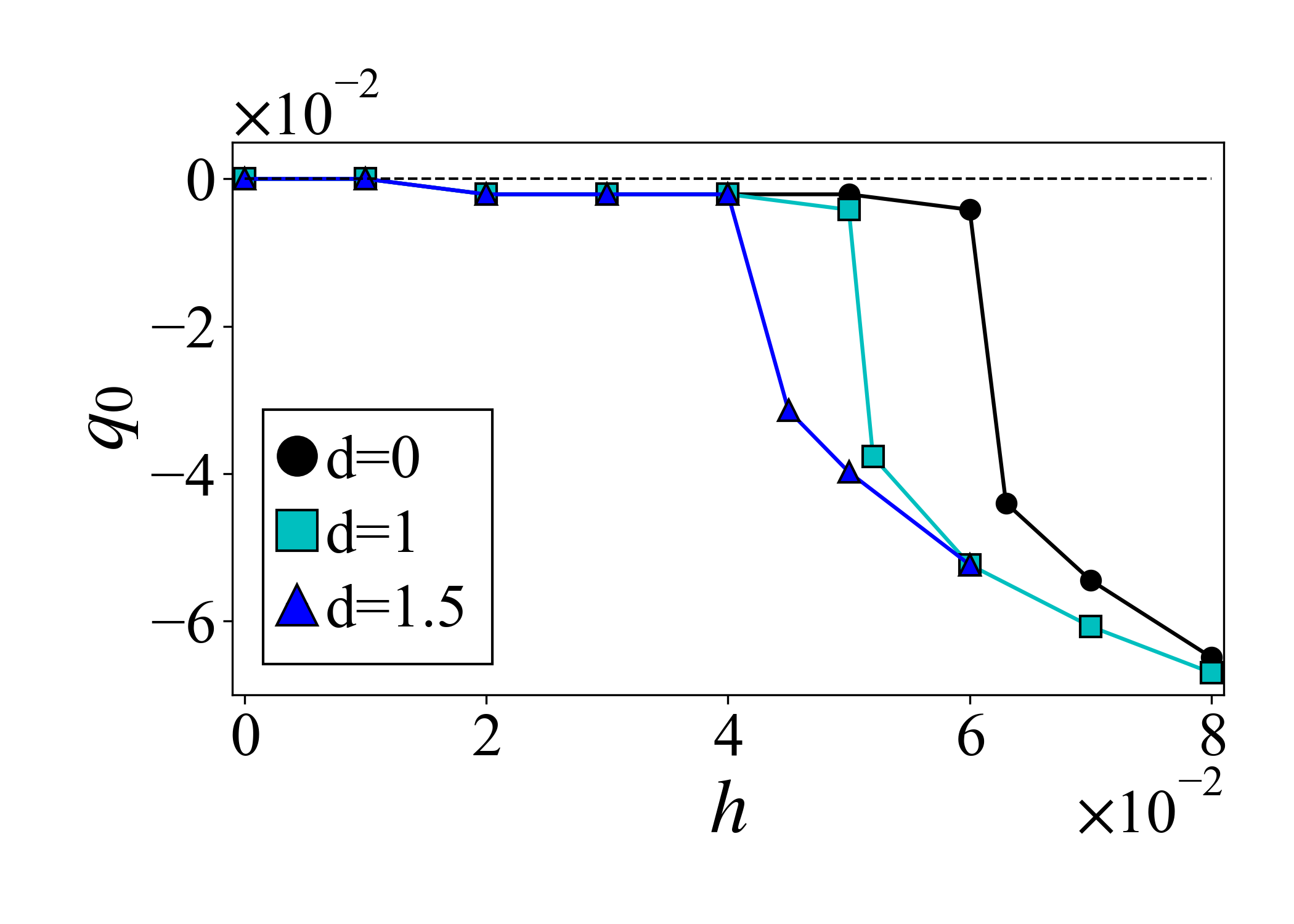}\\
      (c)&&(d)\\
          \includegraphics[width=0.45\textwidth,height=0.25\textheight]{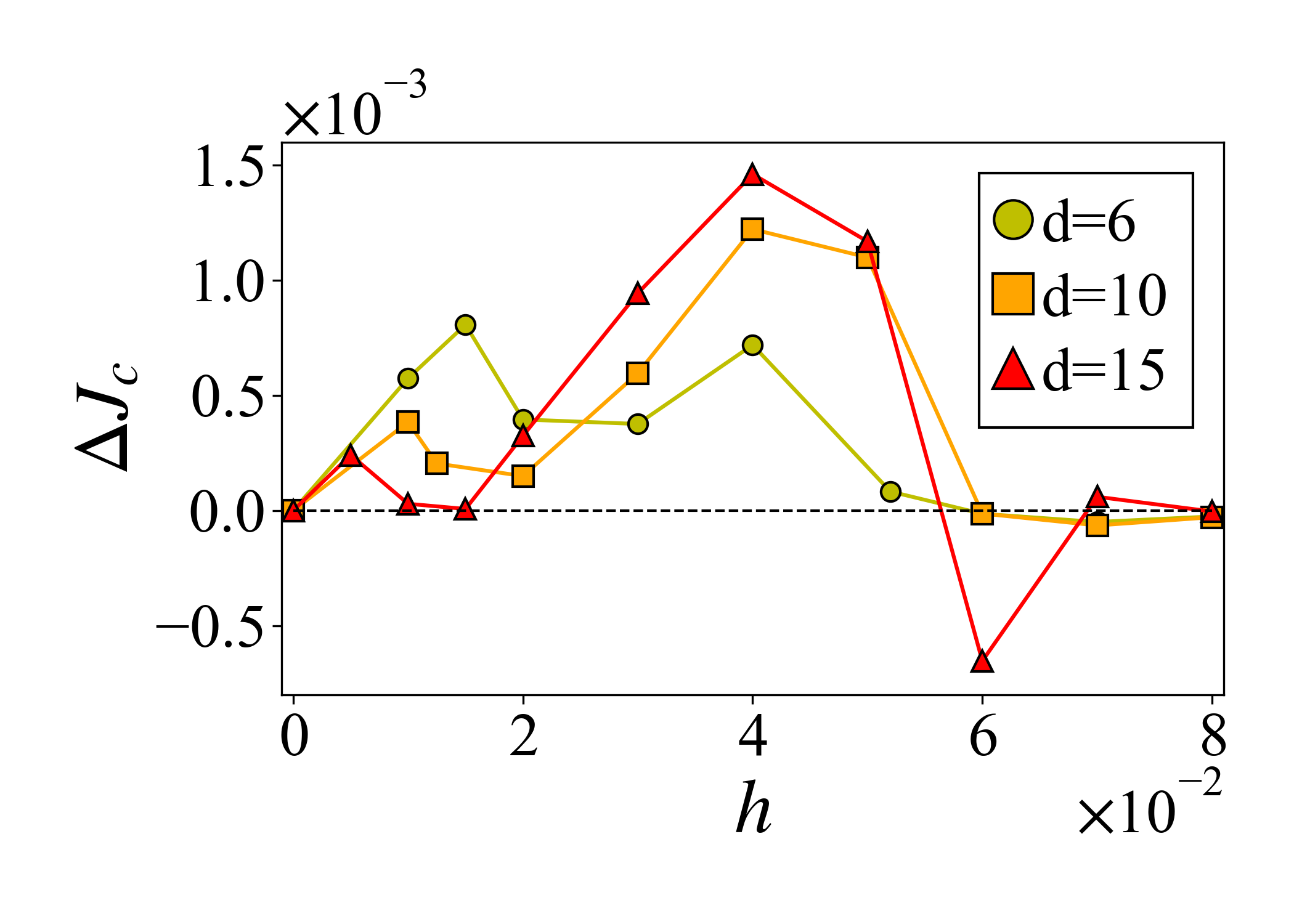}
      &&\includegraphics[width=0.45\textwidth,height=0.25\textheight]{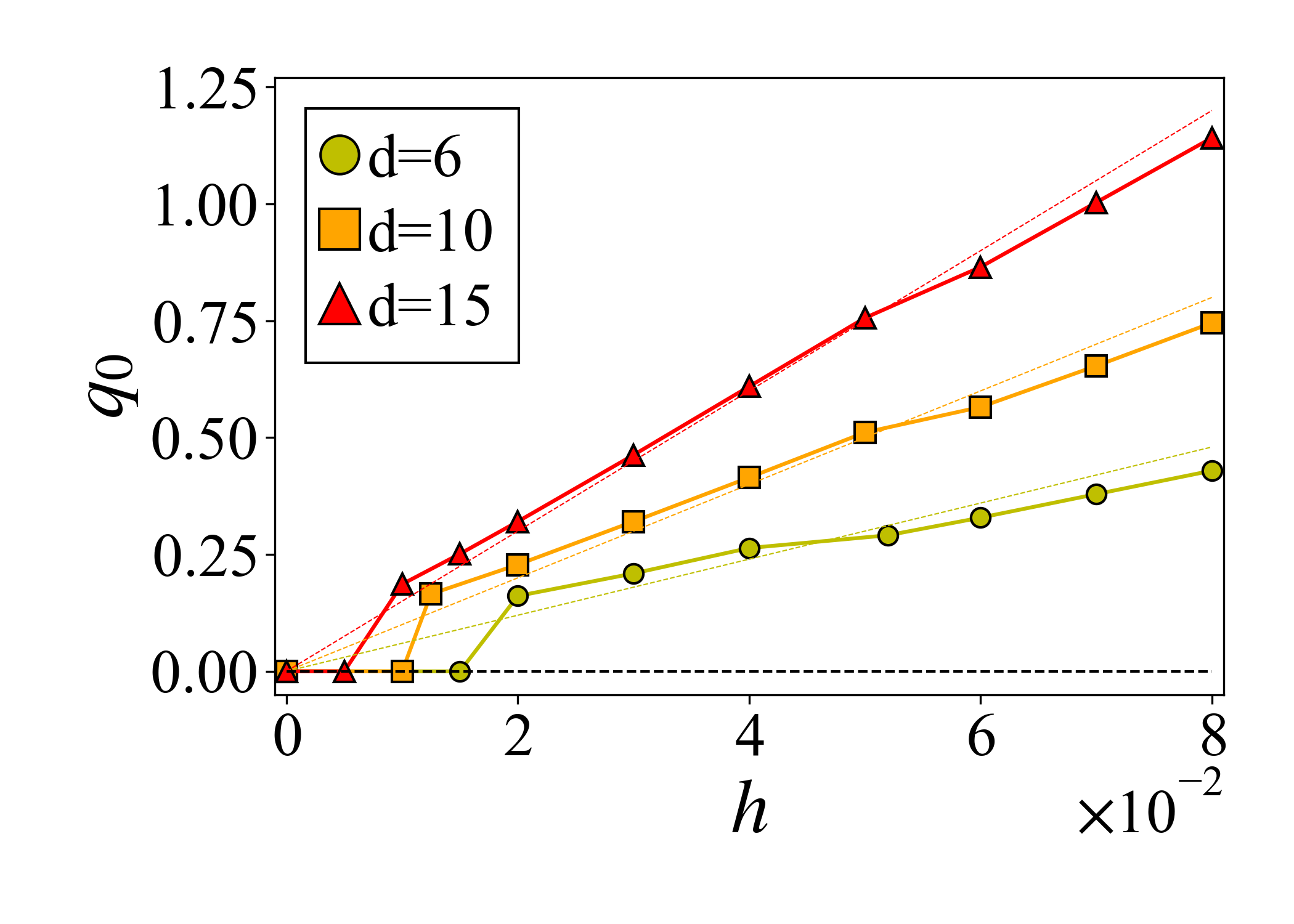}
         \end{tabular}   
    \caption{The magnetic field dependence of [(a), (c)] the nonreciprocity of the critical current $\Delta J_{\rm c}$ and [(b), (d)] the Cooper pair momentum $q_0$ in the equilibrium state. 
    Colored circles, squares, and triangles with guided lines represent $\Delta J_{\rm c}(h)$ and $q_0(h)$ for the parameter of the orbital effect $d=0$, $1$, $1.5$, $6$, $10$, and $15$. 
    The black dashed lines indicate  $J_{\rm c}(h)=0$ and $q_0(h)=0$.
    Colored dashed lines in (d) show $q_0(h)=dh$. %where $d$ corresponds to each color.
    The temperature is set to $T=0.005$, which is much lower than the transition temperature of superconductivity $T_{\rm c} \simeq 0.039$ at the zero magnetic field. 
    Note that we adopt $t_{\perp}=0$ for $d=0$ to reproduce the results for monolayer superconductors~\cite{Daido2022-ox}.
    }
    \label{fig:OE on deltajc and q0}
\end{figure*}

The electric current (the sheet current density) in the $x$ direction is given by
\begin{align}
    j(q)&=j_1(q)+j_2(q) , \label{Eq:current}
\end{align}
\begin{align}
    j_m(q)&=\frac{1}{V}\sum_{\bm k \sigma\sigma'} \partial_{k_x} H_N^m(\bm{k})_{\sigma\sigma'}\,\langle c_{\bm{k}\sigma m}^{\dagger}c_{\bm{k}\sigma' m}\rangle \notag
    \\
    &=\frac{1}{V}\sum_{\bm{k}}\langle\Psi_{qm}^\dagger(\bm{k})\partial_q H_q^m(\bm{k})\Psi_{qm}(\bm{k})\rangle.
\end{align}
The critical current along each direction is defined by
\begin{align}
    &j_{\rm c}(+)\equiv\underset{q}{\mathrm{max}}\,j(q),
\\
%\end{align}
%\begin{align}
    &j_{\rm c}(-)\equiv\underset{q}{\mathrm{min}}\,j(q).
\end{align}
Thereby, the nonreciprocity of the critical current is given by
\begin{align}
    \Delta J_{\rm c}=j_{\rm c}(+)-|j_{\rm c}(-)|.  \label{eq:Delta j_c}
\end{align}

It is helpful to introduce the condensation energy to clarify the superconducting state in equilibrium and its relation to the SDE. 
%whether the superconducting state is helical.
Condensation energy is the difference of free energy between the superconducting and normal states. 
Here, the free energy is calculated by
\begin{align}
    \Omega(\bm{q},\Delta(\bm{q}))&=\frac{1}{2V}\sum_{\bm{k}\sigma m}[[H_N^m(\bm{k})]_{\sigma\sigma}+\Delta_m(\bm{q})^2/U] \notag
    \\
    &\quad -\frac{T}{2V}\sum_{\bm{k}}\mathrm{tr}[\mathrm{ln}(1+e^{-\mathcal{H}_{\bm{q}}(\bm{k})/T)}], 
\end{align}
where $\mathrm{tr}$ represents the trace over the spin and Nambu degrees of freedom.
Thereby, the condensation energy is obtained by
\begin{align}
    F(\bm{q})=\Omega(\bm{q},\Delta(\bm{q}))-\Omega(\bm{q},0), \label{eq:condensed energy}
\end{align}
as a function of the Cooper pair momentum $q$.
By definition, the Cooper pair momentum minimizes the free energy in the equilibrium state, and it is $q_0$ defined by
\begin{align}
    F(q_0)\equiv\underset{q}{\mathrm{min}}\,F(q). \label{eq:q_0}
\end{align}

\section{ORBITAL EFFECT ON THE SDE}
\label{sec:orbital effect on the SDE}

In general, the upper critical field of superconductivity is determined by either the orbital effect, the paramagnetic (Zeeman) effect, or a combination of them. 
In other words, the orbital and paramagnetic effects are the depairing mechanisms due to a magnetic field.
For the intrinsic SDE, the depairing mechanism due to the supercurrent is the kinetic energy of Cooper pairs, which determines the critical current. 
Since the nonreciprocity appears only under the magnetic field in our setup, the interplay of the depairing effects by the magnetic field and supercurrent results in the SDE. 
The orbital effect is therefore expected to impact the SDE, although it was neglected in previous studies for the intrinsic SDE~\cite{Yuan2022-pz,Daido2022-ox,He2022-px,Ilic2022-kh,Daido2022-sd,Ikeda2022-do}.

More specifically, the intrinsic SDE is closely related to the Cooper pair momentum. In our setup, the Cooper pairs get finite momentum because of the deformation of the band structure. The deformation is represented by the shift of electrons' momentum, $\delta\bm{q}_\chi$ [Eq.~\eqref{eq:momentum shift from para}] due to the paramagnetic effect and  $\bm{p}_m$ [Eq.~\eqref{eq:orbital effect}] due to the orbital effect. Therefore, it is naturally expected that not only the paramagnetic effect but also the orbital effect play an essential role in the intrinsic SDE. Note that in a monolayer system the momentum shift $\bm{p}_m$ can be erased by a gauge transformation, and therefore, the orbital effect does not affect any physical phenomena. However, the orbital effect can not be neglected in multilayer systems.

In this section, we show the orbital effect on the nonreciprocal critical current $\Delta J_{\rm c}$ [Eq.~\eqref{eq:Delta j_c}] and the Cooper pair momentum $q_0$ [Eq.~\eqref{eq:q_0}].
As we mentioned above, the orbital effect is parameterized by $d$ [Eq.~\eqref{eq:orbital effect}]. Below we see that the orbital effect on the SDE is qualitatively different between the cases of small and large orbital effects. The criterion is determined by the competition of the paramagnetic effect and orbital effect. 
%In order to briefly understand the results, 
Thus, it is helpful to discuss separately the two situations, whether the orbital effect is smaller or larger than the paramagnetic effect.
The critical value of $d$ depends on the parameters such as a transition temperature of superconductivity. For our choice of parameters, the paramagnetic effect is dominant for $d=0$, $1$, and $1.5$ while the orbital effect is significant for  $d=6$, $10$, and $15$.
In Sec.~\ref{subsec:small orbital effect}, we discuss the small orbital effect on the SDE and the relation to the crossover of helical superconductivity. 
In Sec.~\ref{subsec:large orbital effect}, we discuss the large orbital effect on the SDE and showcase a decoupling transition.

\begin{figure}[htbp]
    \centering
    \begin{tabular}{ll}
    (a)  $h=0.0075$&(b) $h=0.01$\\
        \includegraphics[width=0.23\textwidth,height=0.15\textheight]{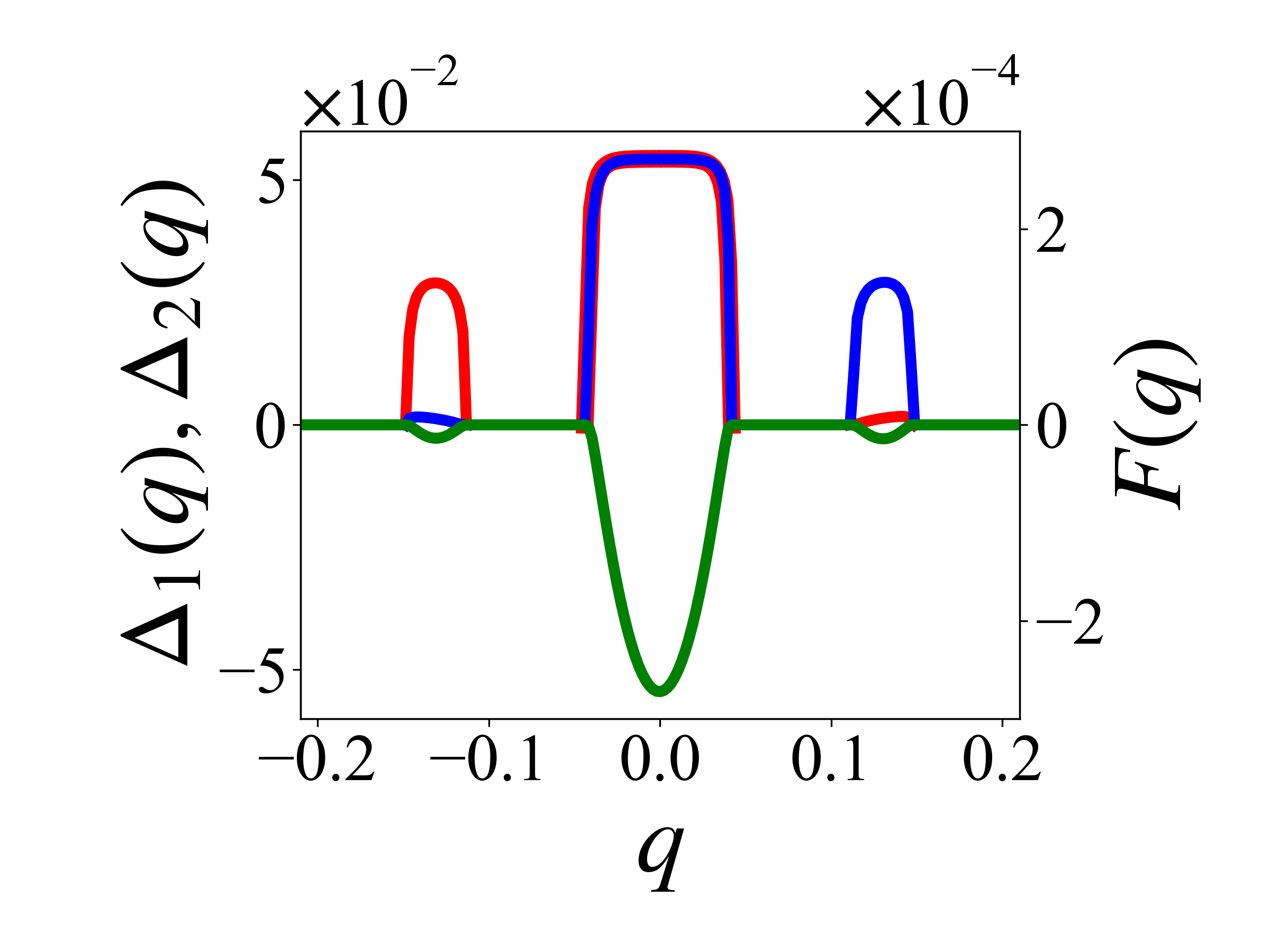}
    &\includegraphics[width=0.23\textwidth,height=0.15\textheight]{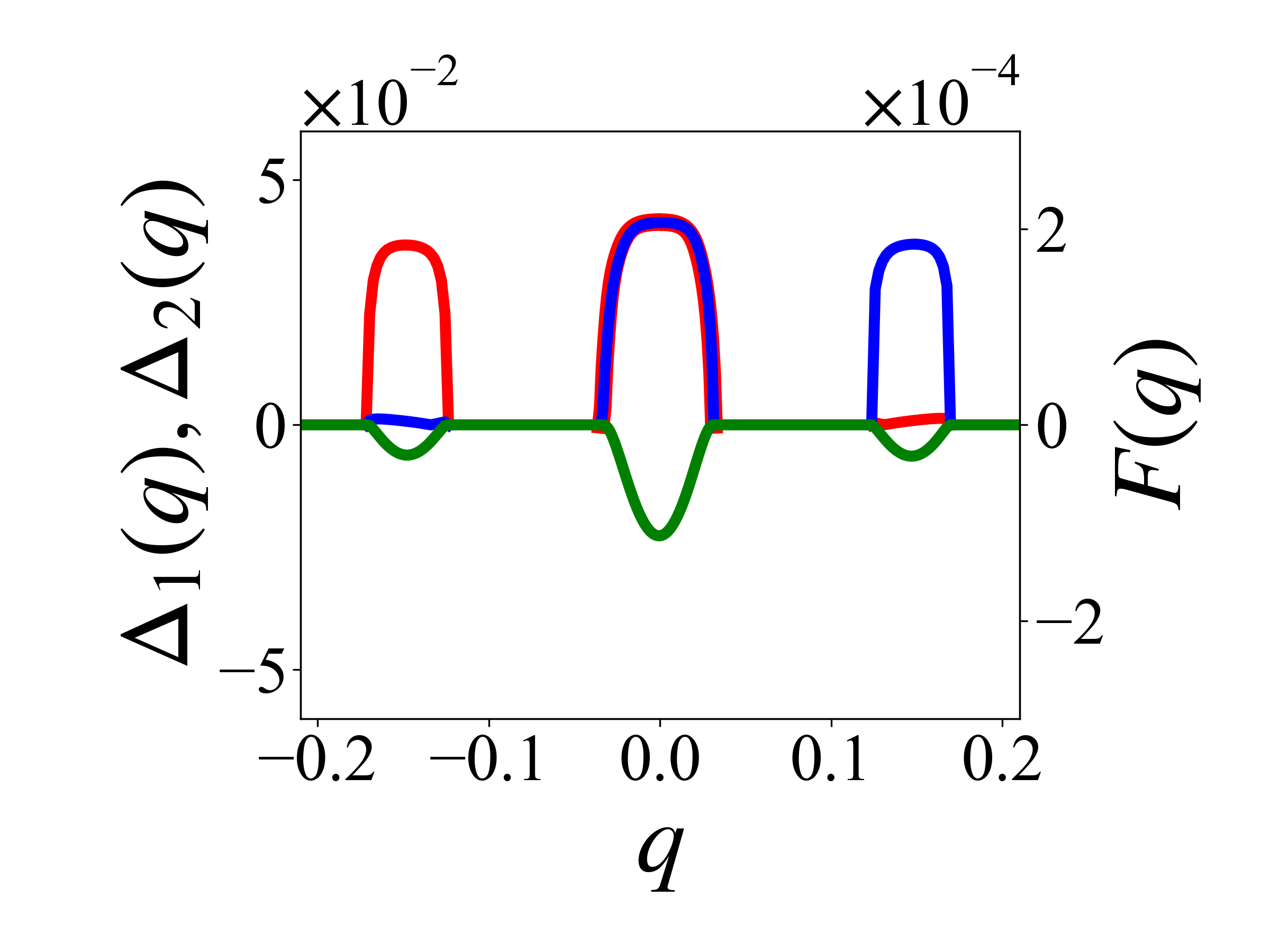}\\
        (c)  $h=0.0125$&(d) $h=0.0135$\\
    \includegraphics[width=0.23\textwidth,height=0.15\textheight]{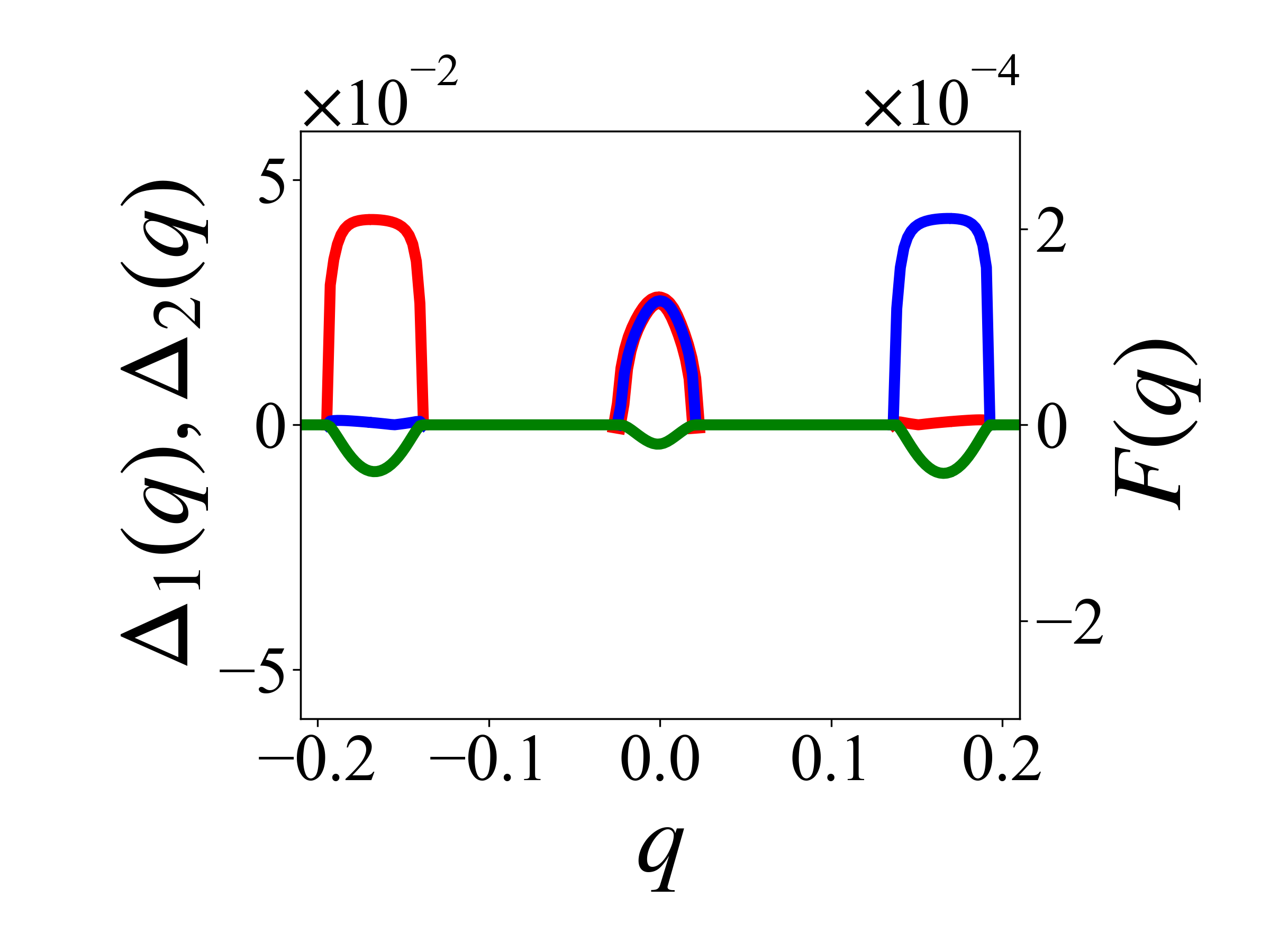}
    &\includegraphics[width=0.23\textwidth,height=0.15\textheight]{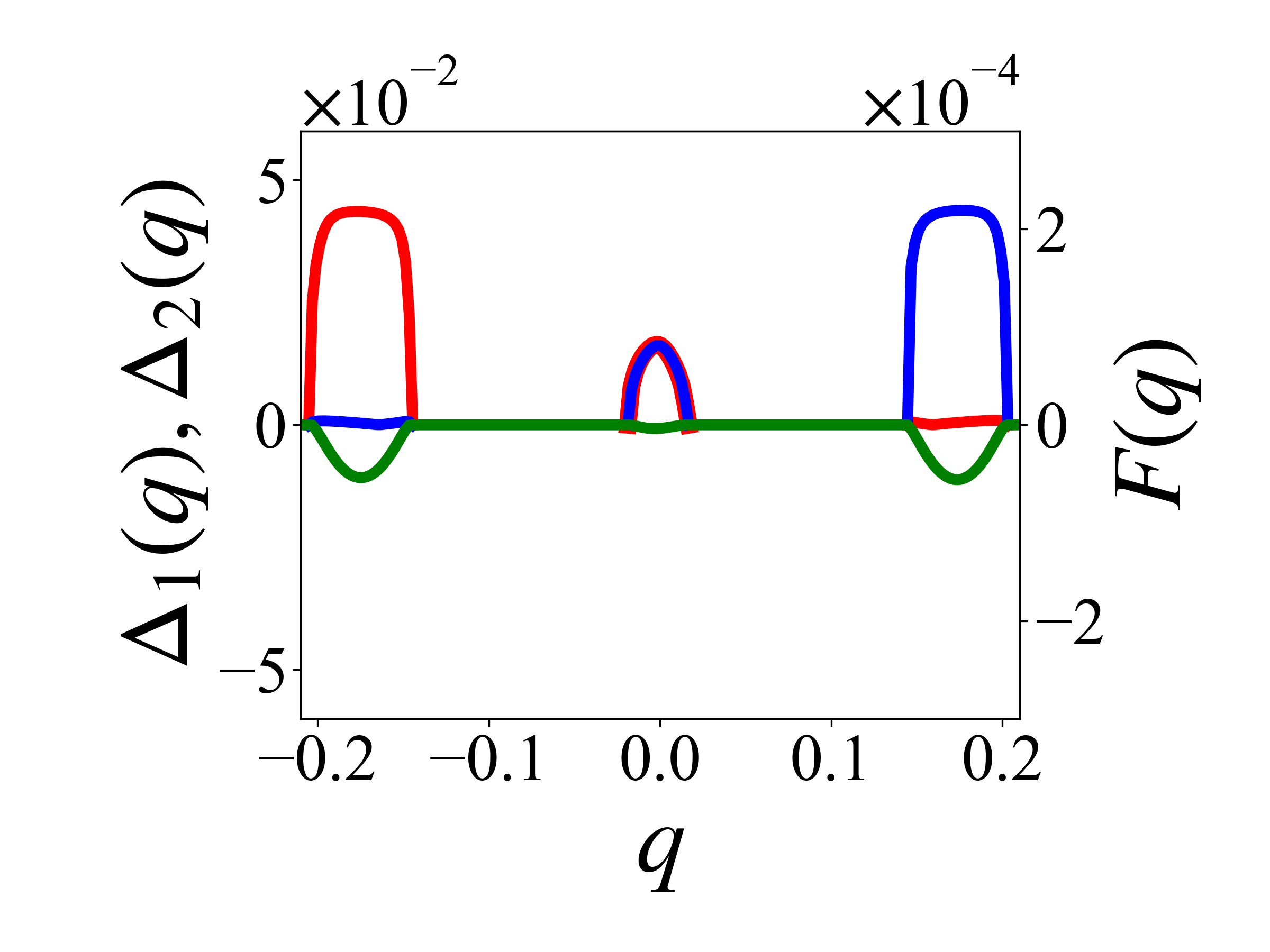}
    \end{tabular}
    \caption{The $q$ dependence of the layer-dependent order parameter, $\Delta_1(q)$ (red lines) and $\Delta_2(q)$ (blue lines), and the condensation energy $F(q)$ (green lines). We assume a large orbital effect with $d=10$ and a low temperature $T=0.005$. The magnetic field is varied as (a) $h=0.0075$, (b) $h=0.01$, (c) $h=0.0125$, and (d) $h=0.0135$.
     %The left and right axis show the energy gap and the condensation energy respectively.
     }
    \label{fig:gap an condensataion}
\end{figure}

\subsection{Small orbital effect on the SDE}
\label{subsec:small orbital effect}

We show in Figs.~\ref{fig:OE on deltajc and q0}(a) and \ref{fig:OE on deltajc and q0}(b) the nonreciprocity of the critical current $\Delta J_{\rm c}$ and the equilibrium Cooper pair momentum $q_0$,  respectively, under a small or vanishing orbital effect. 
The results for $d=0$ correspond to the paramagnetic limit in which the orbital effect is absent. 
By further assuming $t_\perp=0$, the model is equivalent to the monolayer model studied for the intrinsic SDE~\cite{Daido2022-ox}. 
To reproduce the previous results for monolayer superconductors, we set $t_\perp=0$ in the calculation for $d=0$. 
Note that the transition temperatures $T_{\rm c}$ at the zero magnetic field are almost the same between $t_\perp=0$ and $0.1$.
%but the paramagnetic effect, which corresponds to a monolayer system.
In Fig.~\ref{fig:OE on deltajc and q0}(a),
the magnetic field dependence of $\Delta J_{\rm c}$ %and $q_0$ is 
shows qualitatively the same behaviors for $d=0$, $1$, and $1.5$. 
The SDE shows multiple sign changes around a specific magnetic field. 
In monolayer superconductors, the sign change is attributed to the crossover of helical superconductivity at which the equilibrium momentum $q_0$ drastically changes~\cite{Daido2022-ox,Daido2022-sd,Ikeda2022-do}. A comparison of Figs.~\ref{fig:OE on deltajc and q0}(a) and \ref{fig:OE on deltajc and q0}(b) supports this contention even in the presence of the small orbital effect.
Specifically, the sign reversal of SDE and the crossover of helical superconductivity simultaneously occur around $h=0.063$, $0.052$, and $0.045$ for $d=0$, $1$, and $1.5$, respectively, indicating the intact relationship between the SDE and helical superconductivity.
%that is, the oscillation of $\Delta J_{\rm c}$ for the magnetic field $h$ emerges even when there is the small orbital effect.
Now we see that the orbital effect influences the SDE as it decreases the magnetic field where the sign reversal of $\Delta J_{\rm c}$ and the change of $q_0$ take place. 
%Ref.~\onlinecite{Daido2022-ox} suggests the magnitude of the magnetic field at the sign reversal of $\Delta J_{\rm c}$ coincides with that at the crossover of helical superconductivity.
This means that the small orbital effect lowers the crossover field of helical superconductivity.

The origin of the crossover in the helical superconducting state has been studied in a monolayer system, and it is attributed to the paramagnetic depairing effect of Cooper pairs on Rashba-split Fermi surfaces~\cite{Daido2022-ox,Bauer2012-xi,Smidman2017-hb}. 
In bilayer superconductors, not only the paramagnetic effect but also the orbital effect destabilizes the Cooper pairs, and these two effects cooperatively cause the crossover of helical superconductivity. %for the depairing of the Cooper pairs.
Therefore, a small orbital effect accelerates the evolution of $\Delta J_{\rm c}$ and $q_0$ for $h$, resulting in the sign change of the SDE at a lower field.
Consequently, the small orbital effect does not change the behavior of the SDE qualitatively, but it affects the scale of the typical magnetic field.

\subsection{Large orbital effect on the SDE
\label{subsec:large orbital effect}}

Here, we discuss the SDE when the orbital effect is larger than the paramagnetic effect. 
In this subsection, the low-, intermediate-, and high-field regions refer to $0<h<0.02$, $0.02<h<0.05$, and $0.05<h$, respectively.
From the conclusion in Sec.~\ref{subsec:small orbital effect}, we might expect that the large orbital effect further enhances the depairing effect and lowers the magnetic field for the crossover of helical superconductivity.
Contrary to this expectation, we see qualitatively different behaviors of $\Delta J_{\rm c}(h)$ and $q_0(h)$ in Figs.~\ref{fig:OE on deltajc and q0}(c) and \ref{fig:OE on deltajc and q0}(d). 
Figure~\ref{fig:OE on deltajc and q0}(c) shows the two peaks of $\Delta J_{\rm c}(h)$ for any $d$; one in the low-field region and the other in the intermediate-field region. When we further increase the magnetic field, 
the sign reversal of the SDE appears in the high-field region. 
In contrast to the results in Sec.~\ref{subsec:small orbital effect}, 
%On the other hand, in Fig.~\ref{fig:OE on deltajc and q0}(d), 
the equilibrium Cooper pair momentum $q_0$ begins to drastically increase in the low-field region, almost linearly with the magnetic field, as shown in Fig.~\ref{fig:OE on deltajc and q0}(d). 
Thus, the superconducting state drastically changes in the low-field region, and it is different from the crossover of helical superconductivity, as we discuss below.

\begin{figure}[t]
    \centering
    \begin{tabular}{ll}
    (a)  $h=0.04$&(b) $h=0.05$\\
        \includegraphics[width=0.23\textwidth,height=0.15\textheight]{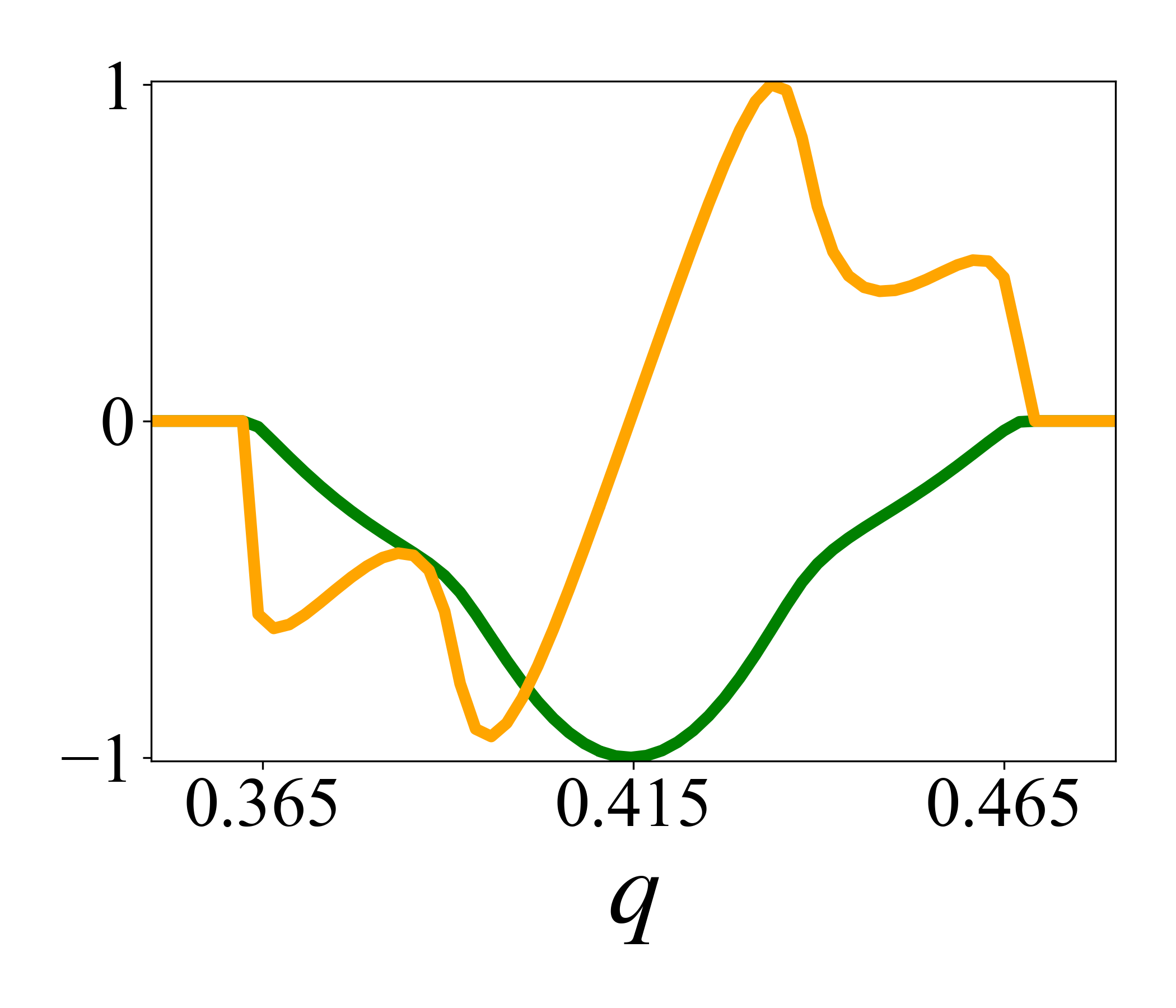}
    &\includegraphics[width=0.23\textwidth,height=0.15\textheight]{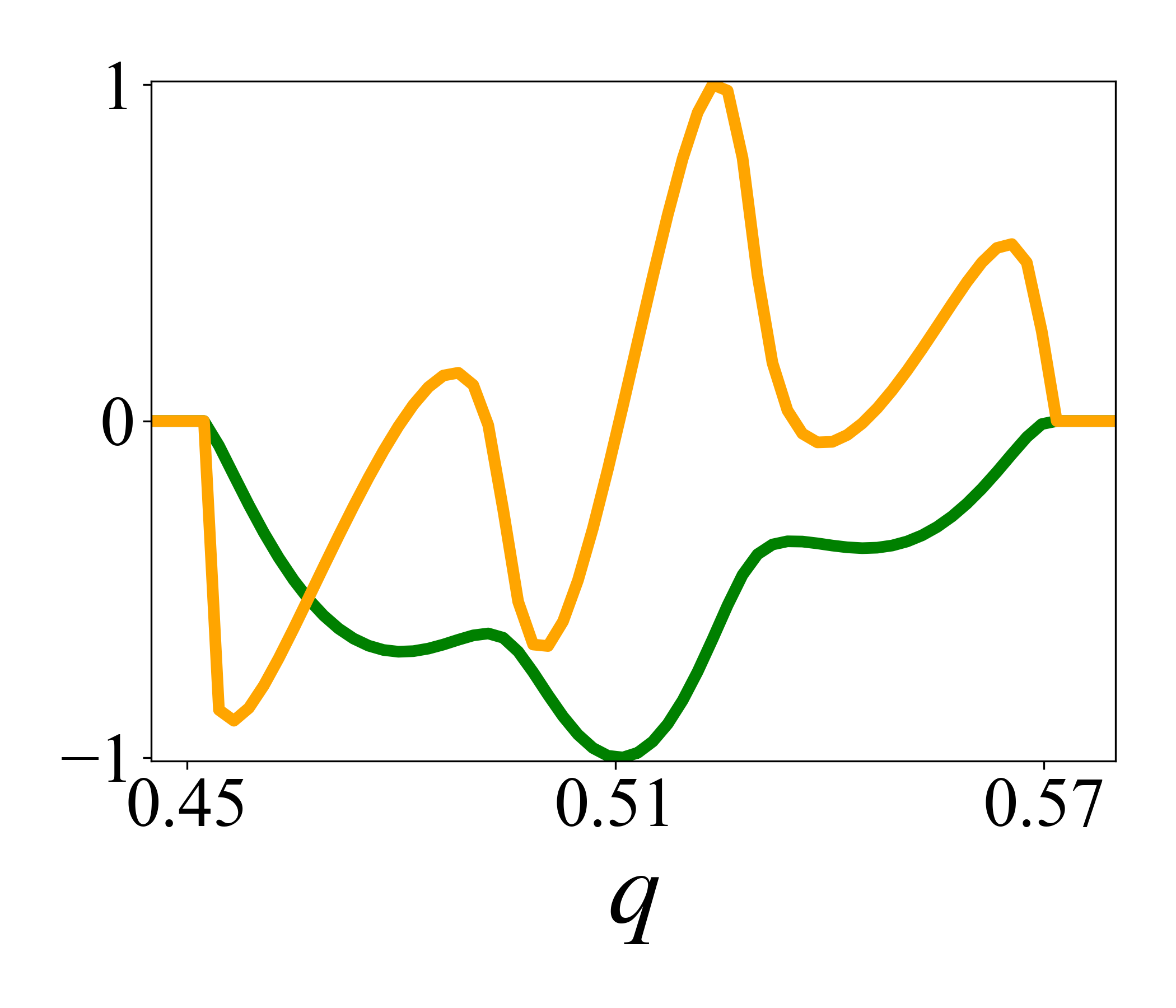}\\
        (c)  $h=0.053$&(d) $h=0.06$\\
    \includegraphics[width=0.23\textwidth,height=0.15\textheight]{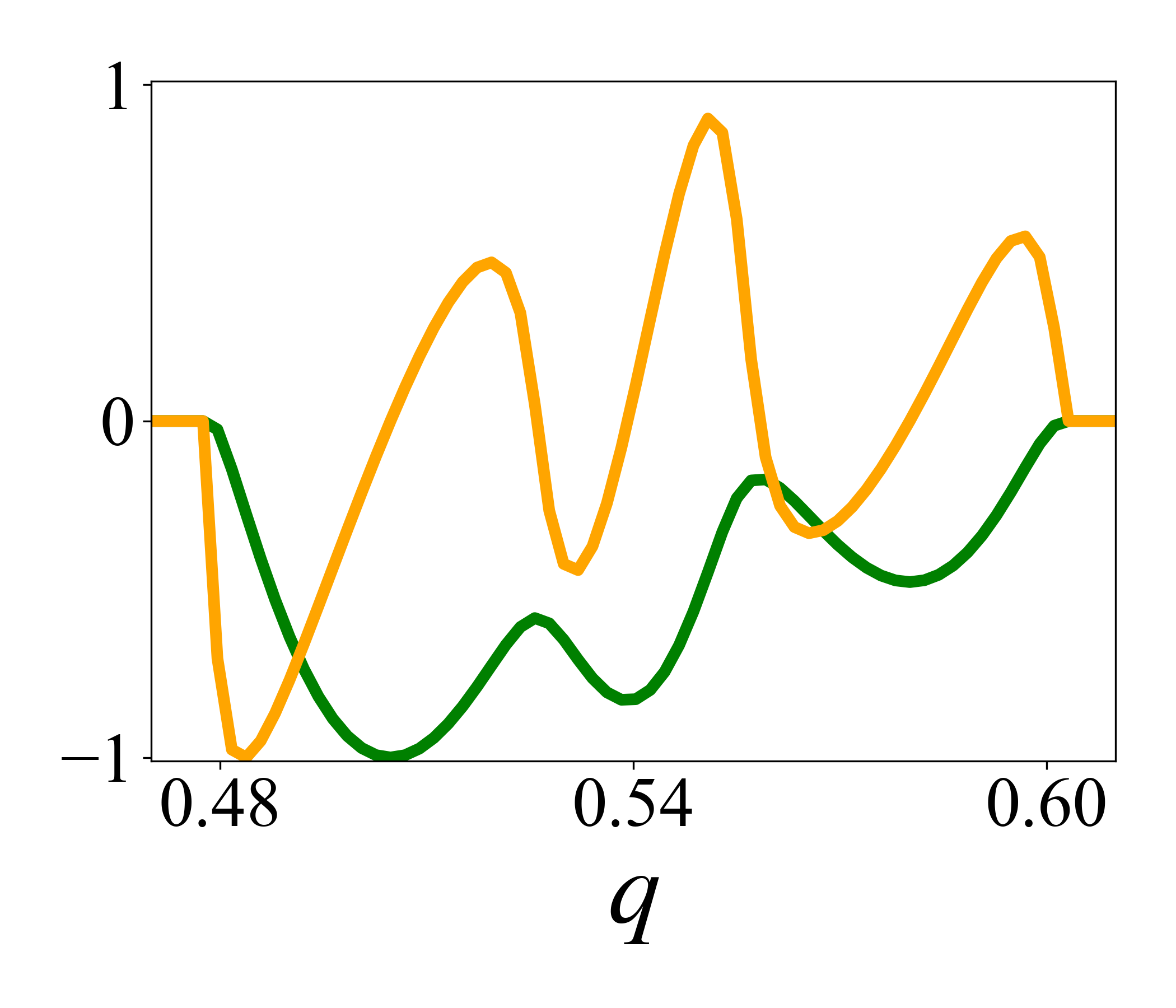}
    &\includegraphics[width=0.23\textwidth,height=0.15\textheight]{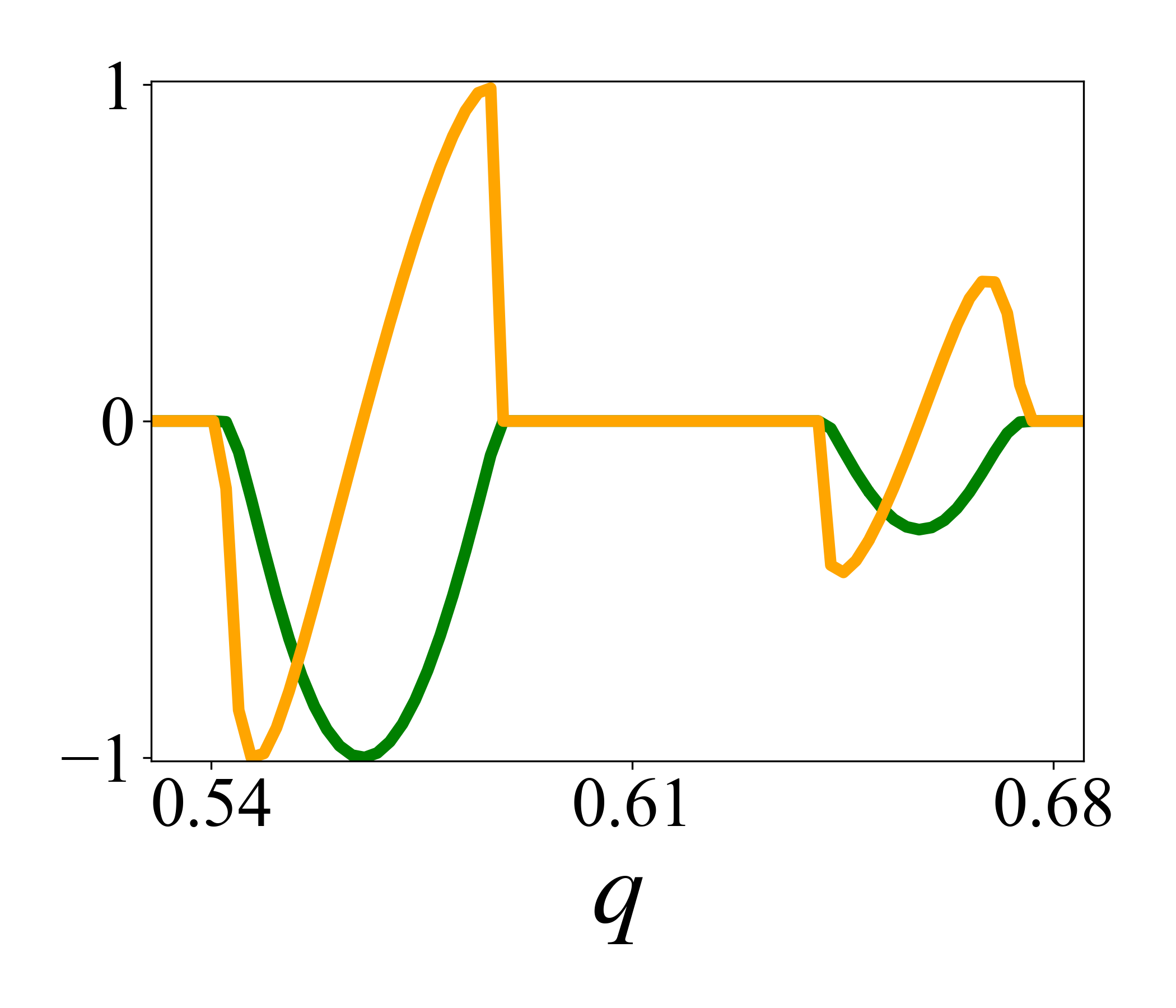}
    \end{tabular}
    \caption{The electric current $j(q)$ (orange lines) and the condensation energy $F(q)$ (green lines) of the decoupled state with $q>0$.
    (a) $h=0.04$, (b) $h=0.05$, (c) $h=0.053$, and (d) $h=0.06$.
    $j(q)$ and $F(q)$ are normalized to $[-1.1]$.
    The parameter of the orbital effect is $d=10$, and the temperature is $T=0.005$.
     }
    \label{fig:j and F}
\end{figure}

To understand the evolution of the superconducting state, it is helpful to show the layer-dependent order parameter $\Delta_m(q)$ [Eq.~\eqref{eq:gap}], the condensation energy $F(q)$ [Eq.~\eqref{eq:condensed energy}], and the electric current $j(q)$ [Eq.~\eqref{Eq:current}] for various values of the magnetic field $h$.
First, we focus on the low-field region.
Figures~\ref{fig:gap an condensataion}(a)-\ref{fig:gap an condensataion}(d) show that 
the condensation energy illustrated by the green lines is negative at $q\sim0$ and $|q| \sim 0.15$, indicating the metastable superconducting states. 
We see that the superconducting states are different between $q\sim0$ and $|q| \sim 0.15$, as revealed by the order parameters $\Delta_m(q)$ illustrated by the red and blue lines.  
In the superconducting state near $q\sim0$, the order parameters of the layers $m=1$ and $2$ are comparable, namely, $\Delta_1(q) \simeq \Delta_2(q)$. The bilayer coupling stabilizes this state, called the coupled superconducting state.
%which refers to the superconducting bilayer coupling.
On the other hand, either $\Delta_1(q)$ or $\Delta_2(q)$ is much larger than the other for the large Cooper pair momentum $|q|>0.1$, representing the bilayer decoupling. 
%consists of either $\Delta_1(q)$ or $\Delta_2(q)$, which represetns the bilayer decoupling.
At zero and low magnetic fields, the coupled superconducting state is naturally more stable than the decoupled superconducting state as seen in Fig.~\ref{fig:gap an condensataion}(a).
The increase in the magnetic field, however, suppresses the coupled superconducting state while it stabilizes the decoupled state as shown in Figs.~\ref{fig:gap an condensataion}(b)-\ref{fig:gap an condensataion}(d).
The strong suppression of the coupled superconducting state is owing to the large orbital depairing effect.
Contrary to the coupled state, an order parameter of superconductivity in either of two layers is suppressed in the decoupled state, and therefore, the orbital depairing effect is almost avoided like in a monolayer superconductor. 
%This is because an order parameter of superconductivity in either of two layers is suppressed and the large orbital effect does not function as the depairing mechanism, but the momentum shift [Eq.~\ref{eq:orbital effect}].
As a result, 
%in Figs.~\ref{fig:gap an condensataion}(c) and \ref{fig:gap an condensataion}(d), 
the stable superconducting state changes from the coupled state to 
the decoupled state as increasing the magnetic field. This is called the decoupling transition.
%is more stable than the coupled state, which can be called a decoupling transition.
As shown in Figs.~\ref{fig:OE on deltajc and q0}(c) and \ref{fig:OE on deltajc and q0}(d), the decoupling transition corresponds to a peak of the nonreciprocal critical current $\Delta J_{\rm c}$ in the low-field region, as it is accompanied by the drastic increase in equilibrium Cooper pair momentum $q_0$.  %in Figs.~\ref{fig:OE on deltajc and q0}(c) and \ref{fig:OE on deltajc and q0}(d).

Next, we discuss the intermediate- and high-field regions.
We show in Fig.~\ref{fig:j and F} how the magnetic field changes the electric current $j(q)$ and the condensation energy $F(q)$ of the decoupled superconducting state.
The development of $j(q)$ and $F(q)$ in Figs.~\ref{fig:j and F}(a)-\ref{fig:j and F}(d) is similar to that in monolayer superconductors~\cite{Daido2022-ox}, which supports an idea that the decoupled state is almost identical to a superconducting state in a monolayer system. 
%that is, $d=0$.
That is why $\Delta J_{\rm c}$ in Fig.~\ref{fig:OE on deltajc and q0}(c) shows the peak in the intermediate-field region and the sign reversal of the SDE in the high-field region.
An essential difference from the superconducting state for $d=0$ is the Cooper pair momentum arising from the orbital effect. 
The momentum shift $|hd/2|$ from Eq.~\eqref{eq:orbital effect} %which is doubled in Cooper pair momentum. 
is much larger than that due to the Rashba-Zeeman effect [Eq.~\eqref{eq:momentum shift from para}]. %is much smaller than that from the orbital effect.
Thereby, the Cooper pair momentum $q_0 \simeq hd$ almost linearly increases with $h$ after the decoupling transition, as shown in Fig.~\ref{fig:OE on deltajc and q0}(d), while $q_0-hd$ shows the crossover behavior similar to Fig.~\ref{fig:OE on deltajc and q0}(b). Note that the shift of Cooper pair momentum is doubled from that of electrons.

At the end of this section, we briefly discuss the relationship between the decoupled superconducting state and the orbital Fulde-Ferrell-Larkin-Ovchinnikov state studied in recent works~\cite{Xie2022-of, Wan_orbital_FFLO, Yuan_orbital_FFLO}.
Reference~\onlinecite{Xie2022-of} has studied a bilayer moir\'{e} system with the Ising SOC and an in-plane magnetic field and suggested that an orbital effect induces a finite momentum pairing phase which is called the orbital Fulde-Ferrell state.
In the orbital Fulde-Ferrell state, Cooper pairs acquire the momentum $q \sim eBd$ with the magnetic field $B$ and the layer separation $d$.
Thus, the momentum agrees with that of the decoupled superconducting state.
We, therefore, expect that the orbital Fulde-Ferrell state results from the decoupling transition. Indeed, the momentum shift by the orbital gauge field and the resulting finite momentum Cooper pairing are ubiquitous as they occur in systems with various symmetries~\cite{watanabe2015,liu2017,nakamura2017}. We expect that the SDE can be a probe of a wide range of finite momentum Cooper pairing states as it is sensitive to the change in the superconducting state. 
Note that Ref.~\onlinecite{Xie2022-of} studied the SDE, but the magnetic field dependence was not shown.

\section{summary and Discussion}
\label{sec:summary and discussion}
In this work, we have revealed the orbital effect on the SDE in a bilayer system with a Rashba spin-orbit coupling and an in-plane magnetic field.
The small orbital effect results in the crossover of helical superconductivity at a lower field, while the large orbital effect induces the decoupling transition accompanied by linearly increasing Cooper pair momentum $q_0$ with the magnetic field. 
We have shown that the SDE drastically changes around the crossover or the transition. Thus, the SDE can be a macroscopic probe of superconductivity with finite momentum Cooper pairs.
Based on these results, we conclude that the orbital effect plays an essential role in the SDE in multilayer heterostructures. 

Interestingly, the SDE shows an oscillating behavior as increasing the magnetic field not only around the crossover of helical superconductivity, as shown in Ref.~\onlinecite{Daido2022-ox}, but also around the decoupling transition. 
Thus, it is expected that the SDE oscillates around the structural transition points of Josephson vortices even in more than three layers of systems. 
Based on this expectation, we can interpret the oscillating behavior of the SDE in the Nb/V/Ta superlattice. Because the oscillation has been observed at the magnetic field much lower than the paramagnetic limiting field~\cite{Kawarazaki2022-sd}, the orbital effect stabilizing the Josephson vortices is more likely to be the origin of the oscillating SDE than the paramagnetic effect.
%When the Josephson vortices undergo successive transitions, 

%Here we remark on another possibility of the sign reversal of $\Delta J_{\rm c}$. In Fig.~\ref{fig:OE on deltajc and q0}(c), there is the bottom in the low-field region after the decoupling transition.
%This bottom approaches the $x$-axis by the increase in $d$.
%Although the mechanism is not unclear, larger $d$ is expected to lead to the sign reversal in the low-field region.

\begin{acknowledgments}
We thank fruitful discussions with Taisei Kitamura, Ryotaro Sano, Jun Oike, Ryo Kawarazaki, and Teruo Ono.  
%This work was supported by JSPS KAKENHI (Grants No. JP18H05227, No. JP18H01178, No. JP19H05825, No. 20H05159, No. 21K13880, and No. 21J14804), JSPS research fellowship, WISE Program MEXT, and SPIRITS 2020 of Kyoto University.
This work was supported by JSPS KAKENHI (Grants Nos.~JP21K13880, JP21K18145, JP22H01181, and JP22H04933).
%\textcolor{blue}{This work was also supported by a Grant-in-Aid for Scientific Research on Innovative Areas “Quantum Liquid Crystals” (KAKENHI Grant No. JP19H05825) from JSPS of Japan.}
\end{acknowledgments}

%\bibliography{ref_pp,ref_for_addition}

%merlin.mbs apsrev4-1.bst 2010-07-25 4.21a (PWD, AO, DPC) hacked
%Control: key (0)
%Control: author (72) initials jnrlst
%Control: editor formatted (1) identically to author
%Control: production of article title (-1) disabled
%Control: page (0) single
%Control: year (1) truncated
%Control: production of eprint (0) enabled

%

%\input{Supplement}

\end{document}